\documentclass[12pt]{article}
\usepackage{a4}
\usepackage{amsfonts}
\usepackage{amssymb}
\usepackage{epsfig}

\newcommand{\be}{\begin{equation}}
\newcommand{\ee}{\end{equation}}
\newcommand{\bea}{\begin{eqnarray}}
\newcommand{\eea}{\end{eqnarray}}
\newcommand{\mbb}{\mathbb}
\newcommand{\ti}{\times}
\newcommand{\half}{\frac{1}{2}}

\newcommand{\mc}{\mathcal}

\begin{document}

\title{
\begin{flushright} \vspace{-2cm}
{\small DAMTP-2005-10, UNH-05-01, UPR-1109-T \\ \vspace{-0.35cm}
hep-th/0502058} \end{flushright}
\vspace{3cm}
{\bf Systematics of Moduli Stabilisation in Calabi-Yau Flux Compactifications}
}
\author{}
\date{}

\maketitle

\begin{center}
V. Balasubramanian\footnote{e-mail:vijay@physics.upenn.edu}$^{,a}$,
P. Berglund\footnote{e-mail:per.berglund@unh.edu}$^{,b}$,
J. P. Conlon\footnote{email:j.p.conlon@damtp.cam.ac.uk}$^{,c}$ and
F. Quevedo\footnote{email:f.quevedo@damtp.cam.ac.uk}$^{,c}$.

\bigskip\medskip\centerline{$^a$ \it David Rittenhouse Laboratories,
  University of Pennsylvania, Philapdelphia, PA 19104, USA.} \smallskip \centerline{$^b$
\it Department of Physics, University of New Hampshire, Durham, NH 03824, USA.}
\smallskip \centerline{$^c$ \it DAMTP, Centre for Mathematical
 Sciences, Wilberforce Road, Cambridge, CB3 0WA, UK.}

\end{center}

\medskip

\begin{abstract}
\noindent

We study the large volume limit of the scalar potential in Calabi-Yau
flux compactifications of type IIB string theory.
Under general
circumstances there exists a limit in which the potential approaches zero from
below, with an associated non-supersymmetric AdS minimum at exponentially
large volume. Both this and its de Sitter uplift are tachyon-free,
thereby fixing all K\"ahler and complex structure moduli.
 Also, for the class of vacua described in this paper, the gravitino mass is independent of
 the flux discretuum, whereas the ratio of the string scale to the 4d Planck scale is hierarchically  small but flux
 dependent.
The inclusion of $\alpha'$ corrections plays a crucial role in the
structure of the potential. We illustrate these ideas through
explicit computations for a particular Calabi-Yau manifold.
\end{abstract}

\thispagestyle{empty}
\clearpage

\tableofcontents

\section{Introduction}
\linespread{1.2}

In \cite{hepth0301240} an elegant scenario was proposed for fixing complex structure,
K\"ahler structure and dilaton moduli in type IIB Calabi-Yau
compactifications. If this scenario is actually
realised in explicit models it will address the main obstacle preventing string theory making contact
with low energy physics. This scenario has already given rise to
many generalisations with applications to realistic models and
cosmological inflation. It has also opened new directions
regarding naturalness issues in string theory in the context of
the string theory landscape. There is some question as to whether such
a low-energy effective potential on the string configuration space
actually exists (see, e.g., \cite{banksrecent}).  However, in this
paper we will not try to address this subtle issue, but simply investigate
the structure of the 
effective potentials resulting from IIB flux compactifications. 

The main ingredients of the KKLT scenario are the presence of fluxes
of RR and NS fields \cite{fluxes,Reviews}, responsible for the fixing of
the dilaton and complex structure moduli, and
non-perturbative effects that fix the K\"ahler moduli. 
(See \cite{fluxpaperstoo} for a sampling of recent work in additional settings.)  
The minimum of the potential can be lifted to a positive value by
additional mechanisms such as adding D-terms through the inclusion of
anti-D3 branes \cite{hepth0301240} or D7-brane magnetic 
fluxes \cite{D7flux}.\footnote{Another possible way to get positive vacuum energy is
  through F-breaking terms in the 
K\"ahler moduli sector, by considering the flux superpotential to be
  of $O(1)$~\cite{hepth0408054}.    Additional ways of  breaking supersymmetry and achieving a metastable de Sitter minimum appear in~\cite{breaking}.}
Model building in this context must contend with several fine-tuning and stability issues. In particular,
the superpotential induced by the fluxes must be
hierarchically small ($<10^{-4}$) in order to obtain solutions
with large volume in which the effective field theory
approximation can be trusted, and
%
the fluxes must fix the dilaton at small string coupling to suppress
loop effects.
%
Although the complex structure moduli and dilaton are
fixed at a minimum of the potential before the non-perturbative and supersymmetry breaking effects are included, these can destabilize some of the scalars (see, e.g., \cite{hepth0411066}).
%
A statistical analysis of the discrete flux choices reveals interesting facts such as a distribution of effective potential extrema that peaks close to conifold points (although this tends to be accompanied by an increase in tachyonic directions in these regions \cite{DouglasStatistics}).  Because of the dependence on the discrete flux choice, the superpotential and the associated supersymmetry breaking scale are scanned by the different vacua, and we are led to regard these quantities as environmentally (rather than dynamically) determined in our world~\cite{hepth0501082}.   Different
proposals have been put forward regarding the preferred supersymmetry breaking scale
\cite{susybreaking}.
In addition to the question of whether such a landscape of vacua
actually exists in a meaningful sense \cite{banksrecent}, it
is clear from the above that the conclusions from this scenario are
highly model dependent. 

  In the present article we will extract model-independent properties of this class of compactifications by studying the large volume limit for a general model with more than one  K\"ahler and complex structure modulus ($h_{12}>h_{11}>1$).
We will argue that the combination of $\left(\alpha'\right)^3$
effects and non-perturbative contributions to the superpotential
will generically give rise to a large
volume non-supersymmetric AdS vacuum, differing from the simplest KKLT
scenario in which the AdS minimum is found to be supersymmetric. To reach this conclusion, we
show that there is a large volume limit in which the potential goes
asymptotically to zero from below, while it is also positive at small
volumes.   This induces the existence of a large volume AdS
minimum.\footnote{While the total volume is very large, some moduli could be stabilized near the string scale, depending
on the particular Calabi-Yau considered.}

The  non-supersymmetric minimum that we find is at exponentially large
volume, and is essentially independent of the value of the flux
superpotential.   We argue that the non-perturbative effects will not
destabilize the flux-stabilized complex structure and dilaton moduli.    
Supersymmetry is broken by the K\"ahler moduli only and the gravitino 
mass is {\it not} flux-dependent: thus it 
does not scan from vacuum to vacuum as the fluxes are tuned, but is
rather peaked at a particular 
value for a particular Calabi-Yau.   The ratio of the string scale
to the 4d Planck scale can be made 
hierarchically small and its value does
depend on the fluxes.   
All of this substantially changes the general picture described above.   
We illustrate this behaviour through a particular model, the
orientifold of $\mbb{P}^4_{[1,1,1,6,9]}$.

\section{Review of type IIB flux compactifications}
\linespread{1.2}

The moduli scalars of string theory compactifications can be stabilized by turning on fluxes on the internal manifold (see \cite{Reviews} for reviews).  We work in IIB theory compactified on Calabi-Yau orientifolds, with RR and NS-NS 3-form fluxes, denoted by $F_3$ and $H_3$ respectively.   These are restricted to have integral cohomology in string theory:
$$
\frac{1}{(2 \pi)^2 \alpha'} \int_{\Sigma_a} F_3 = n_a \in \mbb{Z}, \quad \quad \frac{1}{(2 \pi)^2 \alpha'}
\int_{\Sigma_b} H_3 = m_b \in \mbb{Z},
$$
where $\Sigma_3$ is a 3-cycle in the internal space $M$.  This
framework \cite{hepth0105097} can be viewed as a limit of F-theory
compactifications on an elliptically fibered Calabi-Yau fourfold $X$
\cite{hepth9702165}.  The orientifold 
action results in O3-planes (and wrapped D7-branes/O7-planes) carrying
an anti-D3 brane charge  $\frac{\chi(X)}{24}$ determined by the Euler
characteristic $\chi$ of the fourfold $X$.     
This charge may be cancelled by the 
addition of D3-branes, or through the effective D3-brane charge induced by
$F_3$ and $H_3$ fluxes 
via the the Chern-Simons term in the 10D supergravity action:
\be
S_{CS} \sim \int C_4 \wedge F_3 \wedge H_3.
\ee
This term contributes a D3-brane charge $\int_{M} H_3 \wedge F_3$ leading to the condition
\be
\label{tadpolecancellation}
N_{D3} - N_{\bar{D}3} +  \frac{1}{(2\pi)^4 \alpha'^2} \int H_3 \wedge F_3 = \frac{\chi(X)}{24}.
\ee
Internal magnetic fields on the D7-branes may also contribute D3-brane charge, but we will 
not turn on those fields here.
In order to  avoid the need to stabilize scalars parametrizing D3-brane positions we will require that the fluxes saturate (\ref{tadpolecancellation}).

We will analyze the resulting warped product of a 4d spacetime with an
internal space that is 
conformally Calabi-Yau  \cite{hepth0105097} in the framework of
$\mathcal{N} =1$ supergravity. 
Ignoring gauge sectors, the theory is specified by the K\"ahler
potential and the superpotential.   
The latter takes the form \cite{hepth9906070}
\be
\label{GKVsuperpotential}
W = \int_M G_3 \wedge \Omega,
\ee
where $G_3 = F_3 - \tau H_3$, with $\tau$ being the axion-dilaton
field, and 
$\Omega$ the holomorphic $\left(3,0\right)$ form of the Calabi-Yau.
This does not depend on the K\"ahler moduli. To leading order in $g_s$
and $\alpha'$, the K\"ahler potential is given by the Weil-Petersson
metric derived by Kaluza-Klein reduction of type IIB supergravity,
\be
\label{noscaleK}
K_{no-scale} = -2 \log \left[\mc{V} \right] -
\log\left[-i \int_M \Omega \wedge \bar{\Omega}\right] - \log\left[-i\left(\tau - \bar{\tau}\right)\right],
\ee
where $\cal V$ is the classical volume of $M$ in units of $l_s = (2
\pi) \sqrt{\alpha'}$.
This K\"ahler potential is well-known to possess no-scale structure. Thus, in the
$\mc{N}= 1$ supergravity scalar potential,
\be
\label{Vn=1sugra}
V = e^K \left[G^{i \bar j} D_i W \bar{D}_j \bar{W} - 3 \vert W \vert^2 \right],
\ee
where $i,j$ run over all the moduli,
the sum over K\"ahler moduli cancels the $3\vert W \vert^2$ term, with the resulting potential being
\be
V_{no-scale} = e^K G^{a\bar b}D_a W \bar{D}_b \bar{W},
\ee
where $a$ and $b$ run over dilaton and complex-structure moduli only.
As $V_{no-scale}$ is positive definite, we can locate the
complex structure moduli at a minimum of the potential by solving
\be
\label{csstabilisation}
D_a W = 0 \equiv \partial_a W + (\partial_a K) W = 0.
\ee
This can be done for generic choices of the fluxes and
we denote the value of $W$ following this step as $W_0$.

Since the fluxes are specified by their cohomology, while typical Calabi-Yau manifolds have
$h^3 = \mc{O}(200)$ 3-cycles, there are many discrete 
flux choices that satisfy the consistency condition  (\ref{tadpolecancellation}).
This renders the study of complex structure moduli
stabilization amenable to statistical analysis. This has been carried out
by Douglas and collaborators in a series of beautiful papers \cite{DouglasStatistics}. These analytic
results have been successfully compared
with numerical studies of moduli stabilisation using Monte Carlo simulations to generate
fluxes \cite{hepth0404243,hepth0409215} (see also \cite{hepth0407252}), and the statistical approach has
been extended to other questions \cite{hepth0411061,hepth0411173}. Here we merely summarise these results.

First, the number density of vacua stabilised near a point $z$ in complex structure moduli space is
\be
I_{AD}(z) \sim \left(\frac{\chi}{24}\right)^{(2 h^{2,1} + 2)} \det (- \mc{R} - \omega),
\ee
where $\mc{R}$ is the curvature two-form on complex structure moduli space. This formula is the quantitative
basis for the claim that there exists exponentially many flux vacua. The determinant has a simple form but
a rich structure, for example showing that vacua cluster near conifold loci.

We will make most use of the result that in the discretuum of vacua arising
from flux choices, the values of $e^{K_{cs}} \vert W_0 \vert^2$ are uniformly distributed.
Here $K_{cs}$ refers to the dilaton and complex-structure dependent parts of (\ref{noscaleK}).
This has a similar form to the  gravitino mass $m_{\frac{3}{2}} =
e^{K/2} \vert W \vert$,
but as the full K\"ahler potential $K$ is volume dependent, 
the physical gravitino mass depends on the stabilized value of the volume modulus.

In this article we will study how the leading corrections to the K\"ahler potential and the superpotential, perturbative for the former and non-perturbative for the latter, affect the structure of the scalar potential.
For the features we uncover the subleading corrections 
have subleading effects and can be consistently ignored.

The leading corrections to the K\"ahler potential were computed in \cite{hepth0204254} and
 arise from an $\mc{O}(\alpha'^3)$ term similar to that appearing in
type II compactifications. (See also~\cite{Grimmlouis}.) Measuring dimensionful quantities in units of $l_s = 2\pi
 \sqrt{\alpha'}$, the resulting K\"ahler potential takes
the form \be \label{KahlerPotential} K_{\alpha'} = -2 \log
\left[e^{-3\phi_0 / 2} \mathcal{V} + \frac{\xi}{2}
\left(\frac{-i\left(\tau -
\bar{\tau}\right)}{2}\right)^{3/2}\right] - \log\left[-i \int_M
\Omega \wedge \bar{\Omega}\right] - \log\left[-i\left(\tau -
\bar{\tau}\right)\right], \ee where $\xi = - 
\frac{\zeta\left(3\right) \chi\left(M\right)}{2(2 \pi)^3}$. We
will require $\xi > 0$, which is equivalent to $h^{2,1} > h^{1,1}$: i.e., more complex structure than K\"ahler moduli. This can be
compared with the analogous, exact, correction in pure type II
compactifications \be \label{typeIIcorrection} K = -2 \log \left[
\mc{V} + \frac{\xi}{2} + \textrm{ worldsheet instantons} \right]. \ee 
Although the internal volume is measured in units of $l_s$, the
 $\mc{N} = 1$ supergravity potential is in units of $M_{pl}$. This
 arises from the string theoretic starting point through a standard
 Weyl rescaling to 4-d Einstein frame. 
The dilaton dependence in
(\ref{KahlerPotential}) modifies the K\"ahler metric such that
mixed $G^{\tau \bar{\rho}_i}$ terms no longer vanish. However,
as was shown in~\cite{hepth0408054}, for large volume
 we
note that the behaviour described below is insensitive to the
difference between the corrections in (\ref{KahlerPotential}) and
(\ref{typeIIcorrection}).\footnote{A possible subtlety here is that
 the field strengths that we have turned on could give rise to
 additional complex structure dependent corrections to the K\"ahler
 potential~(\ref{KahlerPotential}). 
Still, by dimensional
 analysis we expect 
that such contributions are relatively suppressed at large volume. (See the discussion section.) We thank Arvind Rajaraman for useful discussions on this point.}

The superpotential is not renormalised at any order in perturbation theory and receives no $\alpha'$ corrections.
However, under certain circumstances it acquires a non-perturbative dependence on
some or all of the K\"ahler moduli through D3-brane instantons
\cite{hepth9604030} or gaugino condensation from wrapped D7-branes
\cite{gc}. It then takes the form
\be
\label{superpotential}
W = \int_M G_3 \wedge \Omega + \sum_i A_i e^{i a_i \rho_i},
\ee
where $A_i$ is a one-loop determinant.   For D3-brane instantons, 
$A_i$ only depends on the complex structure moduli. 
Here  $a_i = \frac{2 \pi}{K}$ with $K \in \mbb{Z}_+$ and $K=1$ for D3-instantons, while
$\rho_i \equiv b_i + i \tau_i$ are the
complexified K\"ahler moduli, with $\tau_i$ the four-cycle modulus, which is the volume of the divisor $D_i\in H_4(M,\mbb{Z})$, given by
\begin{equation}
\tau_i = \partial_{t_i} {\cal V} = {1\over 2} \kappa_{ijk} t^j t^k~.
\label{4to2cycles}
\end{equation}
Here the K\"ahler class is given by $J=\sum_i t^i D_i$ (by Poincare' duality $D_i\in H^2(M,\mbb{Z}$)),
with the $t^i$ measuring the areas of 2-cycles and the classical volume being 
\begin{equation}
{\cal V} = \int_M J^3 = {1\over 6} \kappa_{ijk} t^i t^j t^k~.
\end{equation}
We should understand $\cal V$ as an implicit function of the the complexified 4-cycle moduli $\rho_k$ via the relation between $\tau_i$ and the $t^i$ in~(\ref{4to2cycles}).

Equations (\ref{KahlerPotential}) and (\ref{superpotential}) completely specify the theory.
However, trying to directly visualise the full potential is not illuminating.
We therefore follow KKLT \cite{hepth0301240} and first integrate out
the complex structure moduli. Technically, we stabilise the dilaton and complex structure moduli through solving (\ref{csstabilisation}), and then regard their values as fixed. This leaves a theory only depending on
K\"ahler moduli, which we then stabilise separately. It is important to ask whether the resulting critical point of the full potential
is genuinely a minimum or merely a saddle point. For the simplest implementation of  the KKLT scenario with one
K\"ahler modulus and a rigid Calabi-Yau, the resulting potential has no minima \cite{hepth0411066}.
However, one can argue that true minima can occur in many-modulus models \cite{dd}.
We shall show below that the vacua we find, whether AdS or uplifted dS,
are automatically tachyon-free.

After integrating out the dilaton and complex structure moduli the K\"ahler and superpotential become
\bea
\label{Kalpha'}
K & = & K_{cs}
-2 \log \left[e^{-\frac{3 \phi_0}{2}} \mc{V} +
\frac{\xi}{2} \left(\frac{-i\left(\tau - \bar{\tau}\right)}{2}\right)^{3/2} \right], \\
\label{Wcsfixed}
W & = & W_0 + \sum_n A_n e^{i a_n \rho_n}.
\eea
The dilaton-dependent terms in the K\"ahler moduli dependent part of $K$  are shown as they are
necessary to determine correctly the form of the inverse metric although we then regard the dilaton as fixed by the fluxes.
We note here for subsequent use that as $\mc{V} \to \infty$ the K\"ahler potential behaves as
\be
\label{Klargevol}
e^K \sim \frac{e^{K_{cs}}}{\mc{V}^2} + \mc{O}\left( \frac{1}{\mc{V}^3} \right),
\ee
where all dilaton-dependent terms have been absorbed into $e^{K_{cs}}$.
If we substitute (\ref{Kalpha'}) and (\ref{Wcsfixed}) into equation (\ref{Vn=1sugra}), we obtain the
following potential \cite{hepth0408054}:
\bea
\label{potential}
V & = & e^K \left[ G^{\rho_j \bar{\rho}_k} \left(a_j A_j a_k \bar{A}_k e^{i \left(a_j \rho_j - a_k \bar{\rho}_k\right)}
+ i \left( a_j A_j e^{i a_j \rho_j} \bar{W} \partial_{\bar{\rho}_k} K - a_k \bar{A}_k e^{-i a_k \bar{\rho}_k}
W \partial_{\rho_j} K\right) \right) \right. \nonumber \\
& + & \left. 3 \xi \frac{\left(\xi^2 + 7\xi \mc{V} +
\mc{V}^2\right)}{\left(\mc{V} - \xi\right)\left(2\mc{V} +
\xi\right)^2}
|W|^2\right] \\
& \equiv & V_{np1} + V_{np2} + V_{\alpha'}. \nonumber
\eea

This potential has one well-known class of minima, namely the KKLT
solution \cite{hepth0301240}. This requires a small value of $W_0$
(typically $ < 10^{-4}$), obtained by tuning fluxes, and the
existence of a non-perturbative superpotential depending on all
K\"ahler moduli. The supersymmetry conditions \be D_{\rho_i} W = 0
\ee may then be solved to stabilise all K\"ahler moduli at a
large-volume supersymmetric AdS minimum, which may or may not be a
minimum of the full potential. In \cite{hepth0404257} some
Calabi-Yaus are explicitly constructed for which an appropriate
non-perturbative superpotential will be generated, and a direct
realisation of the KKLT scenario should be possible for the models
described there. There are however two disadvantages to the KKLT
scenario. The first is that one must explicitly check whether a
particular solution is a minimum of the full potential. The second is the small value of
$W_0$ required, given the statistical results on the flux
distribution of $e^K \vert W_0 \vert^2$. Of course, vast numbers
of flux vacua still imply vast-but-not-quite-so-vast numbers of
flux choices with appropriate values of $W_0$. Nonetheless, given
that in the flux `landscape' values of $W_0$ of $\mc{O}(10)$ are
more common than those of order $\mc{O}(10^{-4})$ by a factor
$\sim 10^{10}$, it would be interesting to have examples of
large-volume minima of the potential for large values of $W_0$. We now turn our attention to this.

\section{Large Volume Limit}

In section \ref{GeneralAnalysis} we study the large-volume limit of
the potential (\ref{potential}) 
for a general Calabi-Yau manifold.   
In section~\ref{DouglasModel} we shall then illustrate our ideas through explicit computations on a particular orientifold model.

\subsection{The General Case}
\label{GeneralAnalysis}

The argument for a large-volume AdS minimum of the potential
(\ref{potential}) has two stages. We first
show that there will  in general be
a decompactification direction in moduli space along which
\begin{enumerate}
\item The divisor volumes $\tau_i \equiv \textrm{Im}(\rho_i) \to \infty$.
\item $V < 0$ for large $\mc{V}$, and thus the potential approaches zero from below.
\end{enumerate}
This leads to an argument that there must exist a large volume AdS vacuum.  It might have been expected that the positive
 $(\alpha')^3$ term, scaling as $+\frac{1}{\mc{V}^3}$, will
dominate at large volume
over the non-perturbative terms which are exponentially suppressed.  However, care is needed: the $(\alpha')^3$ term is perturbative in the volume of the entire Calabi-Yau,
whereas the naively suppressed terms are exponential in the
divisor volumes separately.  Hence, in a large volume limit in which some of the divisors are relatively small the non-perturbative terms can compete with the peturbative ones.

The $\left(\alpha'\right)^3$ term in (\ref{potential}) denoted by $V_{\alpha'}$ is easiest to analyse
in the large $\mc{V}$ limit. Owing to the large volume behaviour (\ref{Klargevol})
of the K\"ahler potential, this scales as
\be
\label{Valpha'}
V_{\alpha'} \sim + \frac{3\xi}{16 \mc{V}^3} e^{K_{cs}}|W|^2 + \mc{O}\left(\frac{1}{\mc{V}^4}\right).
\ee
We observe that $V_{\alpha'}$ is always positive and
depends purely on the overall volume $\mc{V}$.

However, $V_{np1}$ and $V_{np2}$ both depend explicitly on the K\"ahler moduli and we must be more precise in
specifying the decompactification limit.
We consider the $\mc{V} \to \infty$ limit in moduli space where $\tau_i \to \infty$ for all moduli
except one, which we denote by $\tau_s$. There are two conditions on $\tau_s$. The first is that this limit
is well-defined; for example, the volume should not become formally negative in this limit. The second
is that $\tau_s$ must appear non-perturbatively in $W$, preferably through D3-instanton effects.
It is however not essential for this purpose
that all K\"ahler moduli appear non-perturbatively in the superpotential.

Let us now study $V_{np1}$ in this limit.
{}From (\ref{potential}),
\be
\label{Vnp1General}
V_{np1} = e^K G^{\rho_j \bar{\rho}_k} \left(a_j A_j a_k \bar{A}_k e^{i \left(a_j \rho_j - a_k \bar{\rho}_k\right)}\right).
\ee
This will be positive definite.
As we have taken $\tau_i$ large for $i \ne s$, the only term not exponentially
suppressed in (\ref{Vnp1General}) is that involving $\rho_s$ alone. $V_{np1}$ then reduces to
\be
V_{np1} =  e^K G^{\rho_s \bar{\rho}_s} a_s^2 |A_s|^2 e^{-2 a_s \tau_s} .
\ee
We need to determine the inverse metric $G^{\rho_i \bar{\rho}_j}$. With $\alpha'$ corrections included, this
is given by (e.g. see \cite{hepth0412239})\footnote{The conventions
  used for the K\"ahler moduli in \cite{hepth0412239} differ
  slightly from ours; as we are in this section only interested in the overall
  sign of the potential these are not important.}
\be
\label{inversemetric}
G^{\rho_i \bar{\rho}_j} = -\frac{2}{9} \left(2\mc{V} + \xi\right) k_{ijk}t^k + \frac{4 \mc{V} - \xi}{\mc{V} - \xi} \tau_i \tau_j.
\ee
In the large $\mc{V}$ limit this becomes
\be
G^{\rho_i \bar{\rho}_j} = -\frac{4}{9} \mc{V} k_{ijk} t^k + 4 \tau_i \tau_j + \textrm{ (terms subleading in $\mc{V}$)}.
\ee
Thus in the limit described above we have
\be
G^{\rho_s \bar{\rho}_s} = -\frac{4}{9} \mc{V} k_{ssk} t^k + \mc{O}(1),
\label{gss}
\ee
with
\be
\label{Vnp1Final}
V_{np1} \sim  \frac{(- k_{ssk}t^k) a_s^2 \vert A_s \vert^2 e^{-2 a_s \tau_s} e^{K_{cs}}}{\mc{V}} +
\mc{O}\left(\frac{e^{-2a_s \tau_s}}{\mc{V}^2} \right).
\ee
Here we have dropped numerical prefactors.   Despite the minus sign in front of (\ref{gss}) this component of the inverse metric will be positive since the K\"ahler metric as a whole is positive definite and this component computes the length squared of the (dual) vector $\partial_{\rho_s} W$.   In the limit we are considering, so long as we remain inside the K\"ahler cone, the leading term which we keep must be positive.

We can perform a similar analysis for $V_{np2}$, from whence negative contributions to the potential arise. We have
\be
V_{np2} = + i e^K 
\left( G^{\rho_j \bar{\rho}_k} a_j A_j e^{i a_j \rho_j} \bar{W} \partial_{\bar{\rho}_k} K - G^{\rho_k \bar{\rho}_j}a_k \bar{A}_k e^{-i a_k \bar{\rho}_k}
W \partial_{\rho_j} K\right).
\ee
The only surviving exponential terms are again those involving $\tau_s$. $V_{np2}$ thus reduces to
\be
V_{np2} = e^K\left[ G^{\rho_s \bar{\rho}_k} \left(i a_s A_s e^{i a_s \rho_s} \bar{W} \partial_{\bar{\rho}_k} K\right) +
G^{\rho_k \bar{\rho}_s}\left(-i a_s \bar{A_s} e^{-ia_s \bar{\rho}_s} W \partial_k K\right)\right].
\label{np2a}
\ee
The form of the inverse metric (\ref{inversemetric}) implies  $G^{\rho_s \bar{\rho}_k} = G^{\rho_k \bar{\rho}_s}$.
The sign of $V_{np2}$ is determined by the value of the axionic field $b_s = \textrm{Re}(\rho_s)$, which
will adjust itself to make $V_{np2}$ negative.    To see this, note
first that at leading order 
in the large volume limit we are considering  $W = W_0 + O(1/\mc{V})$,
so that the only dependence on the
 axion $b_s$ is in $V_{np2}$.    Now write $V_{np2} = e^{ i a_s b_s} X
 + e^{- i a_s b_s} \bar{X}$ where
 we have collected all factors in (\ref{np2a}) except for the axion
 into $X$ and $\bar{X}$.  Extremizing the 
potential with respect to $b_s$, it is easy to see that at a minimum
the axion will arrange its value so as to 
cancel the overall phase from the prefactors to make $V_{np2}$
negative.\footnote{It is interesting to note that the 
phase due to $b_s$ changes sign between the supersymmetric solution, as found by Denef et al~\cite{hepth0404257}, and the non-supersymmetric solution discussed here. This also occurs for the non-supersymmetric vacua of the type discussed in~\cite{hepth0408054} that exist in this model for $W_0=O(1)$ at volumes only slightly larger than the string scale. We thank Kevin Rehberg for making this observation.} 

Thus we may without loss of generality simplify the calculation by replacing
$\rho_s$ by $i \tau_s$ and assuming $A_s$ and $W$ to be both real.
Recall that
$$
\partial_{\rho_k} K = \frac{it^k}{2 \mathcal{V} + \xi} \to \frac{it^k}{2 \mc{V}} + \mc{O}\left(\frac{1}{\mc{V}^2}\right).
$$
Therefore
\be
V_{np2} \sim - e^K a_s A_s W e^{-a_s \tau_s} G^{\rho_s \bar{\rho}_j} \frac{t^j}{\mc{V}} + \mc{O}
\left(\frac{e^{-a_s \tau_s}}{\mc{V}^2}\right).
\ee
We have introduced a minus sign as a reminder that this term will be negative.
Substituting in for $G^{\rho_s \bar{\rho}_k}$ then gives
\bea
V_{np2} & \sim & - e^{K_{cs}} a_s A_s W e^{-a_s \tau_s} \frac{- \frac{4}{9} \mc{V} k_{sjk} t^j t^k  + 4 \tau_s \tau_j t^j}
{\mc{V}^3} + \mc{O}\left(\frac{e^{-a_s \tau_s}}{\mc{V}^3}\right)\nonumber \\
\label{vnp2equation}
& \sim & - e^{K_{cs}} a_s A_s W e^{-a_s \tau_s} \frac{ - \frac{8}{9} \mc{V} \tau_s + 4 \tau_s \tau_j t^j}{\mc{V}^3}
+ \mc{O}\left(\frac{e^{-a_s \tau_s}}{\mc{V}^3}\right)\nonumber.
\eea
{}From the definition of $\tau_i$, the $\tau_s \tau_j t^j$ term will
behave as $\tau_s \mc{V}$ giving a 
uniform volume scaling in the numerator.
Then we conclude that in the limit described above,
\be
\label{Vnp2General}
V_{np2} \sim - \frac{a_s \tau_s e^{-a_s \tau_s}}{\mc{V}^2} \vert A_s W_0 \vert e^{K_{cs}}
+ \mc{O}\left(\frac{e^{-a_s \tau_s}}{\mc{V}^3}\right)\nonumber.
\ee

We may now study the full potential by combining equations (\ref{Valpha'}), (\ref{Vnp1Final}) and
(\ref{Vnp2General}).
\be
\label{fullpotential}
V \sim \left[ \frac{1}{\mc{V}} a_s^2 |A_s|^2 (-k_{ssk}t^k) e^{-2 a_s \tau_s} -
\frac{1}{\mc{V}^2} a_s \tau_s e^{-a_s \tau_s} \vert A_s W \vert
+ \frac{\xi}{\mc{V}^3} |W|^2 \right].
\ee
We have absorbed factors of $e^{K_{cs}}$ into the values of $W$ and $A_s$.
There exists a particular decompactification limit
in which this potential approaches zero from below.
This limit is given by
\be
\label{decomlimit}
\mc{V} \to \infty \quad \quad \textrm{ with } \quad \quad a_s \tau_s = \ln (\mc{V}).
\ee
In this limit the potential takes the following form
\be
\label{volpotential}
V \sim \left[ a_s^2 A_s^2 \frac{(-k_{ssj}t^j)}{\mc{V}^3} - \vert A_s W_0 \vert
\frac{(a_s k_{sjk} t^j t^k)}{\mc{V}^3} + \frac{\xi}{\mc{V}^3}
\vert W_0 \vert ^2 \right] + \mc{O}\left(\frac{1}{\mc{V}^4}\right).
\ee
As in this limit the non-perturbative corrections to $W$
are subleading in $\mc{V}$, we have replaced $W$ by $W_0$.
 We have also written out $\tau_s = \frac{\ln \mc{V}}{a_s}$
in terms of 2-cycle volumes $t^i$.

All terms in equation (\ref{volpotential}) have the same volume dependence
and it is not immediately obvious which is dominant at large volume.
However, the numerator of the
second term of equation (\ref{volpotential}) is quadratic in the 2-cycle volumes,
whereas the others have at most a linear dependence.
As $\tau_i \to \infty$ for all $i$,
all 2-cycles must blow up to infinite volume.
The numerator of the second term is proportional to the volume of the $\tau_s$ 4-cycle  (\ref{4to2cycles}) and thus scales as $\ln \mc{V}$.  Thus, this term scales
as $\frac{\ln \mc{V}}{\mc{V}^3}$, and overcomes the first and third terms which scale schematically
as $\frac{\sqrt{\ln \mc{V}}}{\mc{V}^3}$ and $\frac{1}{\mc{V}^3}$.
In the limit (\ref{decomlimit}) the potential behaves as
\be
V \sim - e^{K_{cs}} \vert A_s W_0 \vert \frac{\ln \mc{V}}{\mc{V}^3},
\ee
and approaches zero from below.\footnote{ 
One concern in the above analysis may be our treatment of $A_s$, which we have treated as a constant.
If the $A_s$ were to depend on the K\"ahler moduli, this may invalidate the argument.
However, we are most interested in superpotentials generated through D3-brane instantons for which
$A_s$ depends only on the complex structure moduli.
In this case we can consistently treat it as constant
when considering solely the K\"ahler moduli. For gaugino condensation the prefactor will
generically depend on all moduli. However, even if $A_s
\propto V^\gamma$, the above argument carries through if we now take $a_s
\tau_s = (\gamma+1) \ln \mc{V}$, although polynomial dependence on the K\"ahler moduli is unlikely to occur
due to the combination of holomorphy and shift symmetry - we thank Liam McAllister for discussions on this point.}

Given this, it is straightforward to argue that there must exist a large volume AdS minimum.
At smaller volumes, the dominant term in the potential ($\ref{fullpotential}$) is either
the non-perturbative term $V_{np1}$ or the $(\alpha')^3$ term $V_{\alpha'}$, depending on the value of
$\tau_s$. Both are positive; the former because the metric on moduli space is positive definite and the
latter because we have required $h^{2,1} > h^{1,1}$.
Thus at small volumes the potential is positive, and so since
the potential must go to zero at infinity and is known to have
negative values at finite volume, 
there must exist a local AdS minimum along the direction in K\"ahler moduli space where the volume changes.   

One may worry that this argument involves the behaviour of the potential at small values of the volume where the
$\alpha'$ expansion cannot be trusted. However, for minima at very large volume, the `small' volumes 
required in the above argument are still large in string units, and we can self-consistently neglect
terms of higher order in $\alpha'$. The relative strength of the $\alpha'$ correction varies between Calabi-Yaus
depending on the precise details of the geometry. For the explicit example studied in the next section, 
the `small' volumes used in the above argument to establish the positivity of the potential 
may be $\mc{O}(10^8)$ in string units.

It remains to argue that the potential also has a minimum in the
remaining directions of the moduli space.   Imagine moving along the
surfaces in the moduli space that are of fixed Calabi-Yau volume, $\mc{V}$.
Then, as one approaches the walls of the K\"ahler cone the first term
in (\ref{fullpotential}) dominates, since it has the fewest powers of
volume in the denominator and since the exponential contributions of
the moduli that are becoming small cannot be neglected.   (Only the
exponential contribution of $\tau_s$ is given in (\ref{fullpotential})
because of the assumed limit, but it is easy to convince oneself that
a similar term will appear for any modulus that is small while the
overall volume is large.)   Thus at large overall volume we expect the
potential to grow in the positive direction towards the walls of the
K\"ahler cone, provided all the moduli appear in the non-perturbative
superpotential.   All told, the potential is negative along the
special direction in moduli space that we have described and
eventually rises to be positive or to vanish in all other directions.   So
we expect an AdS minimum.\footnote{Of course, given the high
  dimensionality of the moduli space, this argument is heuristic.  In
  the next section we will give an explicit example.} 
Since $V\sim O(1/\mc{V}^3)$ at the minimum, while $e^K|W|^2\sim  O(1/\mc{V}^2)$ it is clear that $D_{\rho_s} W \neq 0$ and hence the minimum will be non-supersymmetric

We can heuristically see that the minimum we are arguing for can be at exponentially large volume.  The naive measure of the location of the minimum is the value of the volume at which the second term of
equation (\ref{volpotential}) becomes dominant.  As this occurs when $\ln \mc{V}$ is large, we expect to be able to find vacua at large values of $\ln \mc{V}$.  We will see this explicitly in the example 
studied below.

We can see that the gravitino mass for these vacua will be independent of the flux choice. If the
minimum exists at large volume, it is found by playing off the three terms in equation (\ref{fullpotential})
against each other. If we write $\tilde \mc{V} = \frac{A_s \mc{V}}{W_0}$, then (\ref{fullpotential}) becomes
\be
\label{fullpotential2}
V \sim \left(\frac{A_s^3}{W_0}\right) \left[ \frac{1}{\tilde\mc{V}} a_s^2  (-k_{ssk}t^k) e^{-2 a_s \tau_s} -
\frac{1}{\tilde\mc{V}^2} a_s \tau_s e^{-a_s \tau_s}
+ \frac{\xi}{\tilde\mc{V}^3} \right].
\ee
The minimum of this potential as a function of $\tilde\mc{V}$ is thus independent of $A_s$ and $W_0$ and, given
$a_s$, depends only on the Calabi-Yau. We therefore have
\be
\mc{V} \sim \frac{W_0}{A_s} f(a_s, \mc{M}) + (\textrm{subleading corrections}),
\ee
where $f$ is a function of the geometry. The gravitino mass is then given by
\be
m_{\frac{3}{2}} = e^{\frac{K}{2}}\vert W \vert \approx \frac{A_s}{2 f(a_s, \mc{M})}.
\ee

The number of moduli stabilised depends on the number of
non-perturbative contributions to the superpotential. If all
K\"ahler moduli appear non-perturbatively in the superpotential,
then the dilaton, complex structure and K\"ahler moduli are all
stabilised. If some of the K\"ahler moduli, $\rho_k$,
do not appear in the
superpotential, then as their axionic parts, $b_k$, make no contribution
to the scalar potential, the $b_k$ at least remain unstabilised.
However the volume modulus will be fixed and the
flux-invariant behaviour of the gravitino mass will remain.

Although we have just considered the K\"ahler moduli to find this minimum, it is straightforward to see that
it must actully be a minimum of the full potential.
Reinstating the dilaton-axion and complex structure moduli, this can then be written\cite{hepth0204254}
\bea
\label{Vcsdilaton}
V & = & e^K (G^{a\bar{b}} D_a W \bar{D}_b \bar{W} + G^{\tau \bar{\tau}}
D_\tau W \bar{D}_\tau \bar{W}) + e^K \frac{\xi}{2 \mc{V}} (W
\bar{D}_\tau \bar{W} + \bar{W} D_\tau W) \nonumber \\
& & + V_{\alpha'} + V_{np1} +
V_{np2}.
\eea
Recall that the moduli values found above give rise to a negative value of the potential of
$\mc{O}\left(\frac{1}{\mc{V}^3}\right)$.
The first term in (\ref{Vcsdilaton}) is positive definite and of $\left(\frac{1}{\mc{V}^2}\right)$.
This vanishes iff $D_\tau W = D_{\phi_i} W = 0$. Therefore, any movement of either the dilaton or
complex structure moduli away from their stabilised values would create a positive term
of $\mc{O}\left(\frac{1}{\mc{V}^2}\right)$, which the negative term cannot compete with. Thus this must increase
the potential and therefore the solution above automatically represents a minimum of the full potential.

It is instructive to compare this with what happens for KKLT solutions. The scalar potential is
\be
V = e^{K_{cs}} \Big(\frac{G^{\mu \bar{\nu}} D_\mu W\bar{D}_{\nu} \bar{W}}{\mc{V}^2} -
3\frac{ \vert W \vert^2}{\mc{V}^2}\Big).
\ee
If $D_\mu W = 0$ for all moduli, the potential is negative at $\mc{O}\left(\frac{1}{\mc{V}^2}\right)$.
However, if we move one modulus, for concreteness the dilaton, away from its stabilised value, the resulting
positive definite contribution $e^{K_{cs}} \frac{G^{\tau \bar{\tau}} D_\tau W \bar{D}_\tau \bar{W}}{\mc{V}^2}$
is only of the same order as the minimum. Moving the dilaton alters the value of $e^{K_{cs}}$ and thus may increase
the numerator of the negative term. As the positive and negative contributions are of the same order, we see that
depending on the magnitude of $G^{\tau \bar{\tau}} D_\tau W \bar{D}_\tau \bar{W}$, this may in general decrease
the overall value of the potential. Therefore it is necessary to check explicitly for each choice of fluxes that
the resulting potential has no minimum.

The above solution can be uplifted to a de Sitter vacuum through
the usual mechanisms of adding anti-D3 branes~\cite{hepth0301240}
or turning on magnetic fluxes on D7-branes~\cite{D7flux}. For
concreteness we take the uplift potential to be \be
\label{Vuplift} V_{uplift} = + \frac{\epsilon}{\mc{V}^2}. \ee 
When
$\epsilon = 0$, the above minimum still exists and there are many
values of the moduli for which $V < 0$. For $\epsilon$
sufficiently large, the minimum is entirely wiped out and the
potential is positive for all values of the moduli.   
At a critical value of $\epsilon$ the minimum will pass through zero,
which by construction is still a minimum of the full potential.
Following the arguments in \cite{hepth0301240} it should
be possible to 
tune $\epsilon$ in small steps.     After adding the uplift terms, the
total potential will go 
to zero from above at large volumes because (\ref{Vuplift}) will
overwhelm the $O(1/\mc{V}^3)$ negative terms even in the special limit
that we have been studying.  Hence metastable de Sitter vacua will be
achievable.

\subsection{Explicit calculations for the orientifold of $P^4_{[1,1,1,6,9]}$}
\label{DouglasModel}

We now illustrate the ideas of section (\ref{GeneralAnalysis}) through
explicit
calculations for flux compactifications on an orientifold of
the Calabi-Yau manifold given by the degree 18 hypersurface in
$\mathbb{P}^4_{[1,1,1,6,9]}$. This has been studied by Denef, Douglas and
Florea\cite{hepth0404257} following earlier work in
\cite{hepth9403187}. The defining equation is
\be
\label{definingequation}
z_1^{18} + z_2^{18} + z_3^{18} + z_4^3 + z_5^2 - 18 \psi z_1 z_2 z_3 z_4 z_5 -3 \phi z_1^6 z_2^6 z_3^6
= 0,
\ee
with $h^{1,1}=2$ and $h^{2,1} = 272$.
The complex structure moduli $\psi$ and $\phi$ that have been written in (\ref{definingequation}) are those
two moduli left invariant under the $\Gamma = \mbb{Z}_6 \ti \mbb{Z}_{18}$ action whose quotient gives
the Greene-Plesser construction of the mirror manifold \cite{GreenePlesser}. There are another 270
terms not invariant under $\Gamma$ which have not been written
explicitly, although some will be projected out by the
orientifold action.

We first stabilise the complex structure moduli through an explicit
choice of fluxes. We must solve
\be
\label{StabilisingModuli}
D_\tau W_{cs} = 0 \quad \textrm{ and } \quad D_{\phi_i} W_{cs} = 0~,
\ee
where $W_{cs}$ refers to the flux superpotential,
for the dilaton and complex structure moduli. There are two possibilities.
First, we may of course turn on fluxes along all relevant three-cycles
and solve (\ref{StabilisingModuli}) for all moduli. As we would need
to know all 200-odd periods this is impractical.
We however know of no theoretical reasons not to do it.
The easier approach is to turn on fluxes only along cycles
corresponding $\psi$ and $\phi$, and then solve
\be
\label{SusyStabilising}
D_\tau W_{cs} = D_\psi W_{cs} = D_\phi W_{cs} = 0.
\ee
As described in
\cite{hepth0312104}, the invariance of ($\ref{definingequation}$) under $\Gamma$ then ensures that
$D_{\phi_k} W = 0$ for all other moduli $\phi_k$. The necessary periods have been
computed in \cite{hepth9403187} and appropriate fluxes and solutions
to (\ref{SusyStabilising}) could be found straightforwardly
along the lines of \cite{hepth0404243}\cite{hepth0409215}\cite{hepth0312104}.
This is not our focus in this paper and we henceforth assume this to
have been done.

We now return to the K\"ahler moduli. The K\"ahler geometry is specified by
\be
\mc{V} = \frac{1}{9\sqrt{2}}\left(\tau_5^{\frac{3}{2}} -
\tau_4^{\frac{3}{2}}\right) \nonumber
\ee
$$
\textrm{ with } \tau_4 = \frac{t_1^2}{2} \quad \textrm{ and } \quad
\tau_5 = \frac{\left(t_1 + 6t_5\right)^2}{2}.
$$
Here $\tau_4$ and $\tau_5$ are volumes of the divisors $D_4$ and $D_5$, corresponding to a particular set of 4-cycles, and $t_1$ and $t_5$
2-cycle volumes.
Generally the volume is only an implicit function of $\tau_i$, but here we are
fortunate and have an explicit expression.
As shown in \cite{hepth0404257}, both $D_4$ and $D_5$ correspond to divisors which when lifted to a 
Calabi-Yau fourfold $X$ have
arithmetic genus 1 and thus appear non-perturbatively in the superpotential.
We write this superpotential as
\be
W = W_0 + A_4 e^{i a_4 \rho_4} + A_5 e^{i a_5 \rho_5}.
\ee
We now take the limit described in section \ref{GeneralAnalysis},
in which $\mc{V} \to \infty$ (and hence $\tau_5\to\infty$) and $\tau_4\sim \log{\mc{V}}$. Note that the
alternative limit $\tau_4 \to \infty$ with $\tau_5\sim \log{\mc{V}}$ would not be
well-defined, as the volume of the
Calabi-Yau becomes formally negative.

\begin{figure}[ht]
\linespread{0.2}
\begin{center}
\epsfxsize=0.85\hsize \epsfbox{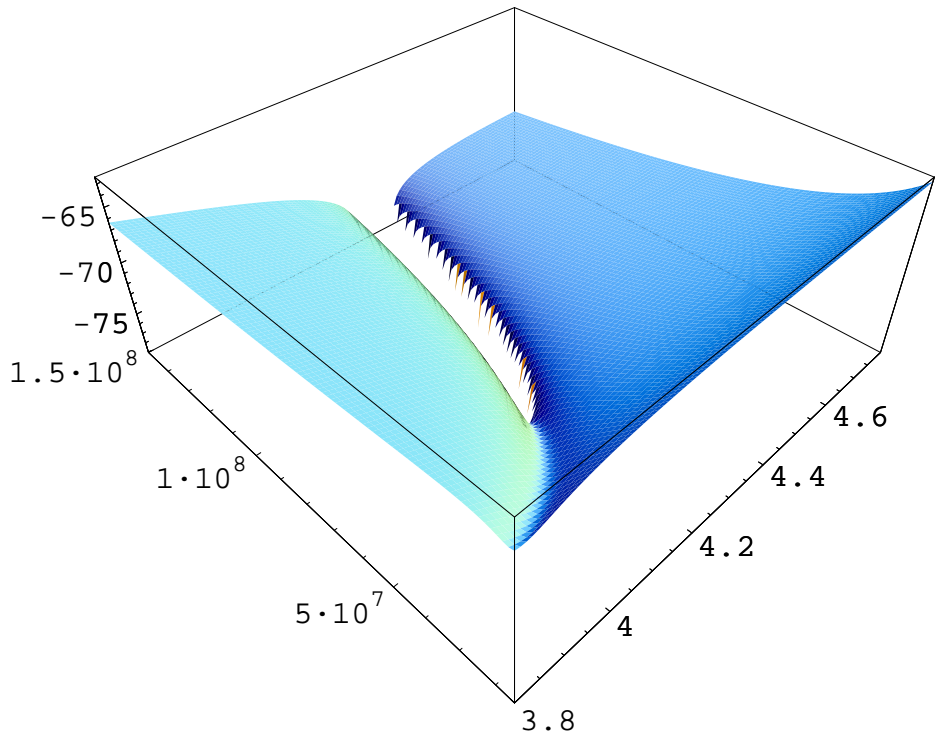}
\end{center}
\caption{ $\ln (V)$ for $P^4_{[1,1,1,6,9]}$ in the large volume
  limit, as a function of the divisors $\tau_4$ and $\tau_5$. The void
  channel corresponds to the region 
  where $V$ becomes negative and $\ln (V)$ undefined. As $V \to 0$ at
  infinite volume, this immediately shows that a large-volume minimum
  must exist. Here the values $W_0 = 20,
  A_4 = 1$ and $a_4 = 2\pi$ have been used.}
\label{LargeVolumeLimit}
\end{figure}

The $\alpha'$ correction is given by equation (\ref{Valpha'}). For
$V_{np1}$ and $V_{np2}$ we must compute the inverse metric, which in
this limit is given by
\bea
G^{\rho_4 \bar{\rho_4}} & = & 24 \sqrt{2} \sqrt{\tau_4} \mc{V} \sim \sqrt{\tau_4} \mc{V}, \\
G^{\rho_4 \bar{\rho_5}} = G^{\rho_5 \bar{\rho_4}} & = & 4 \tau_4 \tau_5 \sim \tau_4 \mc{V}^\frac{2}{3}, \\
G^{\rho_5 \bar{\rho_5}} & = & \frac{4}{3} \tau_5^2 \sim \mc{V}^\frac{4}{3}.
\eea
We can then compute $V_{np1}$ and $V_{np2}$ with the result that the
full potential takes the schematic form
\be
\label{examplepotential}
V \sim \left[ \frac{1}{\mc{V}} a_4^2 |A_4|^2 \sqrt{\tau_4} e^{-2 a_4 \tau_4} -
\frac{1}{\mc{V}^2} a_4 \tau_4 e^{-a_4 \tau_4} \vert A_4 W \vert
+ \frac{\xi}{\mc{V}^3} |W|^2 \right].
\ee
where numerical coefficients have been dropped.  We have implicitly extremized with respect to the axion $b_4$ to get a negative sign in front of the second term as described below equation (\ref{np2a}).
It is obvious that in the limit
\be
\tau_5 \to \infty \quad \textrm{ with } \quad a_4 \tau_4 = \ln \mc{V},
\ee
the potential approaches zero from below as the middle term of equation (\ref{examplepotential})
dominates. This is illustrated in figure \ref{LargeVolumeLimit} where we plot the
numerical values of $\ln (V)$.

The location and properties of the AdS minimum may be found analytically.
To capture the form of equation (\ref{examplepotential}), we write
\be
\label{Vgeneral}
V = \frac{\lambda \sqrt{\tau_4} e^{-2 a_4 \tau_4}}{\mc{V}}
- \frac{\mu}{\mc{V}^2} \tau_4 e^{-a_4 \tau_4} + \frac{\nu}{\mc{V}^3}.
\ee
The axion field $b_5$ has been ignored as terms in which it appears are
exponentially suppressed.

We regard $V = V(\mc{V}, \tau_4)$, and solve
$$
\frac{\partial V}{\partial \mc{V}} = \frac{\partial V}{\partial \tau_4} = 0.
$$
One may easily check that the first of these equations may be
rearranged into a quadratic and solved for $\mc{V}$ to give
\be
\label{Vsoln}
\mc{V} = \frac{\mu}{\lambda} \sqrt{\tau_4} e^{a_4 \tau_4} \left(1 \pm
\sqrt{1 - \frac{3 \nu \lambda}{\mu^2 \tau_4 ^{\frac{3}{2}}}} \right)
\ee
We also have
\be
\frac{\partial V}{\partial \tau_4} = 0 \Rightarrow
\frac{\lambda \mc{V} e^{-a_4 \tau_4}}{\tau_4^{\half}} \left(\half - 2
a_4 \tau_4 \right) - \mu \left(1 - a_4 \tau_4 \right) = 0.
\ee
We then use (\ref{Vsoln}) to obtain an implicit equation for $\tau_4$,
\be
\label{tausoln}
\left(1 \pm \sqrt{1 - \frac{3 \nu \lambda}{\mu \tau_4^{\frac{3}{2}}}} \right)\left( \half - 2 a_4 \tau_4 \right)
= (1 - a_4 \tau_4).
\ee
We do not need to solve this fully; as we require
$a_4 \tau_4 >> 1$ to be able to ignore higher instanton corrections, we can use this
to simplify (\ref{tausoln}) and solve for $\tau_4$ and $\mc{V}$, obtaining
\bea
\tau_4 & = & \left( \frac{4 \nu \lambda}{\mu^2} \right)^{\frac{2}{3}},
\nonumber \\
\mc{V} & = & \frac{\mu}{2 \lambda} \left( \frac{4 \nu \lambda}{\mu^2}
\right)^{\frac{1}{3}}
e^{a_4 \left( \frac{4 \nu \lambda}{\mu^2} \right)^{\frac{2}{3}}}.
\eea

For the potential of (\ref{Vgeneral}),
$$
\lambda \sim a_4^2 \vert A_4 \vert ^2, \quad \mu \sim a_4 \vert A_4
W_0 \vert, \quad \textrm{ and }
\nu \sim \xi \vert W_0 \vert^2.
$$
We then have
\be
\label{TauVolFinal}
\tau_4 \sim (4 \xi)^{\frac{2}{3}} \quad \textrm{ and }
\mc{V} \sim \frac{\xi^{\frac{1}{3}} \vert W_0 \vert}{a_4 A_4} e^{a_4 \tau_4}.
\ee
This formula justifies our earlier claim that these vacua can
generically be at exponentially large volume. 

For the $P^4_{[1,1,1,6,9]}$ example,
$$
\xi = 1.31, \quad \lambda = 3\sqrt{2} a_4^2 \vert A_4 \vert^2, \quad
\mu = \half a_4 \vert A_4 W_0 \vert, \quad \nu = 0.123 \vert W_0
\vert^2.
$$
\bea
\label{examplesolutions}
\Rightarrow \tau_4 & \approx & 4.11 + \textrm{ (small terms)}, \nonumber \\
\mc{V} & \approx & 0.12 \frac{W_0}{a_4 A_4} \ti e^{4.11 a_4} + \textrm{ (small terms)}.
\eea
These analytic results agree well with the exact locations of the
minima found numerically, with the small error almost entirely
due to the approximation made in solving equation (\ref{tausoln}).
As discussed in section \ref{GeneralAnalysis},
the values of the K\"ahler moduli found above combine with the flux-stabilised complex structure moduli
to give a minimum of the full potential (\ref{Vn=1sugra}).

We note that both the overall and divisor volumes are clearly larger than the string scale, and that
 as long as $a_4$ is not too small
the gravitino mass  $e^{K/2} \vert W \vert$ is well below the Planck
 scale. For $a_4 = 2\pi$ (as it is for D3-brane instantons), $A_4 = 1$, $W_0 = 10$, we obtain $\mc{V}
 \approx 3 \ti 10^{10}$ in string units.
The gravitino mass $m_{\frac{3}{2}} = e^{K/2} W$ is then given by $3
 \ti 10^{-10} M_{p} = 4 \ti 10^9$ GeV and the string scale
 $M_s =  (M_p g_s)/ \sqrt{\mc{V}} \sim 7 \ti 10^{12}$ GeV for $g_s =
 \frac{1}{10}$.  
Here $M_p$ is the 4d Planck scale and we take
$M_p \sim
10^{19} \, \textrm{GeV}$.
So we have an explicit
 realisation of the intermediate scale string scenario \cite{intermediate}. As discussed below,
 this is independent of the flux-induced value of $W_0$.  
 
\section{Discussion}

We have shown that there exists a decompactification direction in moduli space along
which nonperturbative D3-instanton effects dominate over 
the perturbative $\alpha'^3$ corrections and therefore a large volume
minimum is induced in a very general class of 
compactifications (for which $h_{12}>h_{11}>1$).\footnote{In the special large volume limit that we have described, the potential will go to zero from below even if $h_{12} < h_{11}$, namely even if the parameter $\xi$ in our analysis is negative.  However, in this case the small volume behavior is harder to analyze, and so it is not clear if the non-supersymmetric minimum that we have found occurs.} 
There are several very interesting features about the above limit
which may have important implications:

\begin{enumerate}
 \item 
As exemplified by equation (\ref{TauVolFinal}), the mechanism described
here results in internal spaces that are exponentially large in string units. 
The largest such volumes arise for the case $a_4 = 2
\pi$, when the non-perturbative
dependence on $\tau_4$ arises through D3-brane instantons. However,
gaugino condensation with $a_4 = \frac{2 \pi}{N}$ can also lead to
large volume vacua. 
Having exponentially large volume $\mc{V}$ implies a realization of the large
extra dimensions scenario in which the fundamental string scale is
hierarchically smaller than the Planck scale since the string scale $M_s$ and the 4d Planck scale are related by $M_s \sim M_p/\sqrt{\mc{V}}$.
The ratio of the string scale to the 4d Planck scale
is scanned over the different vacua in the sense
that it depends explicitly on the fluxes through the stabilized
Calabi-Yau volume. For the particular example discussed here we find
an intermediate string scale $M_s \sim 10^{12}\ $ GeV, which has been
claimed to have some phenomenological virtues \cite{intermediate}.
\item
The gravitino mass is given by $m_{\frac{3}{2}}^2 = e^K \vert W
 \vert^2$. As $e^K \sim \frac{1}{\mc{V}^2}$ and $\mc{V} \propto W_0$,
this is unaffected by the value of $W_0$ arising from the flux choice.
This gives the striking result that the discretuum of gravitini masses
should be sharply peaked around one particular value.
For the $P^4_{[1,1,1,6,9]}$ case, we have
$$
m_{\frac{3}{2}} \approx 4 \times 10^9\  \textrm{GeV}.
$$
Given the universality of the gravitino mass across the space of flux
choices, it would be interesting to know whether this might also
be universal or near-universal across the space of Calabi-Yau manifolds. This 
might result in the possibility of well-defined physical predictions from
the discretuum of flux vacua.
\item
The volume obtained scales as $\mc{V} \propto W_0$. Thus increasing
$W_0$ increases the internal volume. This behaviour is opposite
to that encountered in the KKLT scenario, where small
values of $W_0$ are required for large volumes.
It is also evident from equation (\ref{TauVolFinal}) that in the case of
D3-brane instantons with $a_4 = 2\pi$ essentially all values of $W_0$
will give an acceptable solution.
Note that the results of Douglas and Denef \cite{DouglasStatistics} show that
$e^{K_{cs}} \vert W_0 \vert^2 $ is uniformly distributed in the
discretuum of flux choices and thus `typical' choices of flux give rise
to large values of $W_0$.
\item 
The mechanism described above relies on there being at least two K\"ahler moduli, which is of course the generic case.
If all K\"ahler moduli appear non-perturbatively in the
superpotential, then the mechanism above will stabilise all
moduli. 
However, even if only some moduli appear in the
superpotential, then as argued in section (\ref{GeneralAnalysis}) the
volume modulus will still be stabilised at exponentially large volume,
and we also expect the flux-invariance of the gravitino mass to be unaltered. 
Further,  as argued at the end of section \ref{GeneralAnalysis}, the stabilised
values of the moduli will automatically represent a minimum of the
full potential. Thus there is no need to perform the almost impossible check
that all hundred-odd directions in moduli space are non-tachyonic.

\item The tuning is, in a precise sense, minimal. That is, the only tuning performed
is that necessary to ensure that the dilaton is at weak
coupling. This will always be necessary until strong-coupling string
dynamics is better understood. In particular, there is no need to tune
the value of $W_0$ - as emphasised above, any value will do.

\end{enumerate}

We also note that as the solutions described above are not tachyonic,
the increase in tachyons found near the conifold locus by Denef and
Douglas \cite{DouglasStatistics} would not be applicable here.
Therefore  the natural accumulation of solutions around the conifold locus
indicated by the Ashok-Douglas formula and checked in 
\cite{hepth0404243, hepth0409215} for particular cases, now holds for our solutions.

Finally,  we should consider effects that may destabilise the
above behaviour. Higher order $\alpha'$ corrections are subleading in
volume and so will not alter the above behaviour. More important would
be other $\alpha'^3$ corrections. The $\alpha'^3$ correction used here
descends from the $\mc{R}^4$ term in the ten-dimensional action, the coefficient of
which is a modular form \cite{hepth9808061}. The coefficient
$\xi$ should then be promoted to a dilaton-dependent modular form. As
we fix the dilaton using fluxes, this will not affect the qualitative
behaviour described above although it may affect the numerical values.
There may also be other $\alpha'^3$ corrections descending from the
flux terms in ten dimensions.\footnote{We thank A. Sinha and M. Green
  for interesting conversations about this point.} Little is known
about these extra terms in this
setting and it would be interesting to understand their effects.~\footnote{A preliminary estimate of these terms (e.g. the term
  proportional to $G_3^8$) shows 
them to scale as $1/\mc{V}^k$ with $k>3$ in their contributions to the scalar potential.
  Thus they are suppressed compared to the terms we have considered.
  A detailed study of the 
effects of these extra terms lies beyond the scope of this paper.  We thank A. Sinha for discussions on this point.}
In the presence of extra $1/{\cal V}^3$ corrections, the quantitative
behaviour of our solution will change but the qualitative behaviour
will remain.
It is still an interesting open question to determine their exact effect.
There is also the general concern about loop corrections after
supersymmetry breaking. In principle these will be suppressed by
powers of the dilaton field that can be always tuned to be small by
the choice of fluxes, but a better understanding of these corrections
would be worthwhile. Warping effects should be suppressed at large
volume, but so are the $\alpha'$ corrections and it would be interesting
to study their relative magnitudes. Finally, $\alpha'$ corrections due to
localised sources should be considered. 

We have worked in a particular limit of F-theory which can be
interpreted as a type IIB orientifold. It would be of interest to
address these questions in more
general F-theory settings, which include D-brane moduli\cite{hepth0501139}. 

We have argued that, at least to leading order, supersymmetry is
broken by the K\"ahler moduli and unbroken by the complex structure
moduli. It should be possible to calculate the soft susy-breaking
terms engendered by this minimum. Another interesting direction worth exploring is the study of
this class of potentials in the context of brane/antibrane
\cite{bantib} or closed moduli (racetrack) inflation
\cite{racetrack}. 

We hope to address some of
these issues in the future.

\section*{Acknowledgments}
PB would like to thank A. Rajaraman and K. Rehberg for
discussions and the organizers of the Conformal Field Theory 2nd
Reunion Conference at Lake Arrowhead for a stimulating environment
where some of this work was done. PB, JC and FQ thank the organizers
of the String Vacuum Workshop at the Max-Planck Institute, 
Munich, for hosting an exciting conference where this project was
initiated. JC and FQ are grateful to J.J. Blanco-Pillado,
C.P. Burgess,
J. Cline,  
D. Cremades, C. Escoda,  M. G\'omez-Reino, M. Green, L. Ib\'a\~nez, L. McAllister, A. Sinha and
K. Suruliz for discussions.  
VB is supported in part by the DOE under grant DE-FG02-95ER40893, by
the NSF under grant PHY-0331728 and by an NSF Focused Research Grant
DMS0139799. PB is supported by NSF grant PHY-0355074 
and by funds from the College of Engineering and Physical Sciences  at
the University of New Hampshire.
JC thanks EPSRC for a research studentship. FQ is partially supported by 
PPARC and a Royal Society Wolfson award.

\end{document}